\documentclass[referee]{jfm}
\usepackage{color}
\usepackage{amstext}
\usepackage{amssymb}
\usepackage{graphicx}
\usepackage{amsmath}

\newcommand\const{\mathrm{const}}

\newcommand\vX{\boldsymbol{X}}

\newcommand\vF{\boldsymbol{F}}

\newcommand\vV{\boldsymbol{V}}

\newcommand\va{\boldsymbol{a}}
\newcommand\vb{\boldsymbol{b}}
\newcommand\vc{\boldsymbol{c}}

\newcommand\vk{\boldsymbol{k}}

\newcommand\vl{\boldsymbol{l}}
\newcommand\vm{\boldsymbol{m}}

\newcommand\vx{\boldsymbol{x}}

\newcommand\vq{\boldsymbol{q}}
\newcommand\vQ{\boldsymbol{Q}}

\newcommand\vPi{\boldsymbol{\Pi}}

\begin{document}

{\title[Theory of a triangular micro-robot] {Theory of a triangular micro-robot}}

\author[V. A. Vladimirov]
{V.\ns A.\ns V\ls l\ls a\ls d\ls i\ls m\ls i\ls r\ls o\ls v}

\affiliation{Dept of Mathematics, University of York, Heslington, York, YO10 5DD, UK}

\pubyear{2010} \volume{xx} \pagerange{xx-xx}
\date{October 2nd 2012}

\setcounter{page}{1}\maketitle \thispagestyle{empty}

\begin{abstract}

In this paper we study the self-propulsion of a triangular micro-robot (or $\triangle$-robot) which consists
of three spheres connected by three rods; the rods' lengths are changing independently and periodically.
Using the asymptotic procedure containing the two-timing method and distinguished limit arguments, we obtain
analytic expressions for  self-propulsion velocity  the angular velocity.  Our calculations show that a
$\triangle$-robot rotates with  constant angular velocity around its centroid, while the centroid moves in a
circle. The important special case of zero angular velocity represents rectilinear translational
self-propulsion with constant velocity.

\end{abstract}

\section{Introduction \label{sect01}}

The studies of micro-robots represent a flourishing modern research topic which strives to create a
fundamental base for modern applications in medicine and technology, see \emph{e.g.}
\cite{Purcell, Koelher, NG+, Dreyfus, Yeomans1, Paunov, Lefebvre,  Gilbert, Golestanian, Golestanian1, Yeomans,
Pietro, Lauga, Romanczuk}. The simplicity of micro-robot's geometry represents the major advantage in
contrast to extreme complexity of self-swimming microorganisms,
\emph{e.g.} \cite{PedKes, VladPedl, Pedley, Polin}. This advantage allows us to describe the motion of micro-robots
in greater depth.

In this paper, we study the self-propulsion of a triangular three-sphere micro-robot (we call it
$\triangle$-robot), see the figure. We formulate the problem for a $\triangle$-robot with arbitrary sides,
but the final expressions are presented only for the case when all mean sides are equal. Our calculations
show that the $\triangle$-robot rotates with  constant angular velocity $\Omega$ around its centroid, while
the centroid itself moves along a circle with the same angular velocity $\Omega$.  The important special case
of $\Omega=0$ represents rectilinear translational self-propulsion with constant velocities. To obtain these
results we employ the two-timing method and distinguished limit arguments, which lead to a simple and
rigorous analytical procedure. Our approach is technically different from all previous methods employed in
the studies of micro-robots (except
\cite{VladimirovX3,VladimirovX4}). The possibility to describe explicitly any motion of a $\triangle$-robot
shows the strength of our method. The used version of the two-timing method has been developed in
\cite{Vladimirov0,Vladimirov1,VladimirovMHD}.
\begin{figure}
\centering\includegraphics[scale=0.7]{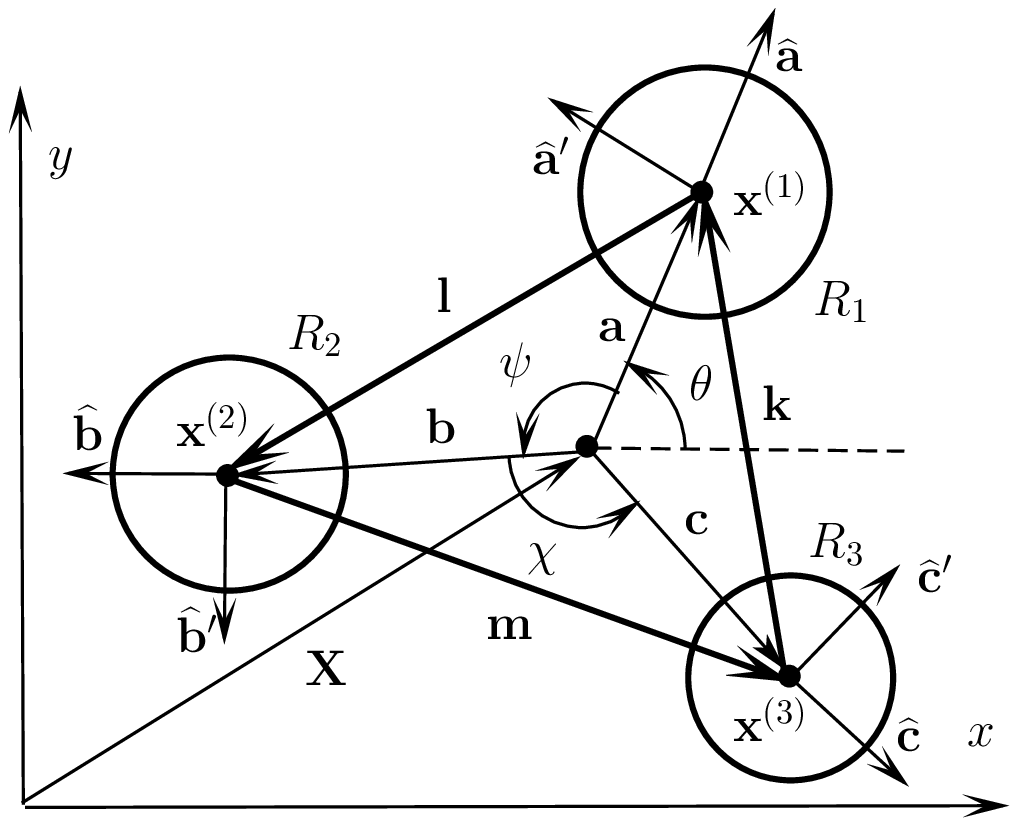}
\label{3-robot}
\end{figure}

\section{Formulation of problem\label{sect01}}
We consider the $\triangle$-robot consisting of three rigid spheres of equal radii $R$ connected by three
rods of different lengths $k$, $l$, and $m$. In plane cartesian coordinates $(x,y)$ the centers of the
spheres are $\vx^{(\nu)}=(x_1^{(\nu)},x_2^{(\nu)})$, $\nu=1,2,3$ and the triangle (with vertices in centres
of spheres) is composed of vectors $\vk\equiv \vx^{(1)}-\vx^{(3)}$, $\vl\equiv \vx^{(2)}-\vx^{(1)}$, and
$\vm\equiv
\vx^{(3)}-\vx^{(2)}$, see the figure. We describe the centers of the spheres as
$$
\vx^{(1)}=\vX+\va,\quad\vx^{(2)}=\vX+\vb,\quad \vx^{(3)}=\vX+\vc
$$
where $\vX=(X,Y)$ is a centroid radius-vector such that vectors $\va$, $\vb$, $\vc$, unit vectors
$\widehat{\va}$,  $\widehat{\vb}$, $\widehat{\vc}$, and angles $\theta$, $\psi$, $\chi$ are introduced as
\begin{eqnarray}
&&\va+\vb+\vc=0;\quad \vk=\va-\vc,\ \vl=\vb-\va,\ \vm=\vc-\vb;\label{abc}\\
&&\va=a\,\widehat{\va},\quad \vb=b\,\widehat{\vb},\quad\vc=c\,\widehat{\vc}\label{abc-units}\\
&&\widehat{{\va}}\equiv
\left(
\begin{array}{c}
    \cos\theta  \\
     \sin\theta
  \end{array}
\right),\quad
\widehat{{\vb}}\equiv
\left(
\begin{array}{c}
    \cos(\theta+\psi)  \\
     \sin(\theta+\psi)
  \end{array}
\right),\quad
\widehat{{\vc}}\equiv
\left(
\begin{array}{c}
    \cos(\theta+\psi+\chi)  \\
     \sin(\theta+\psi+\chi)
  \end{array}
\right)\nonumber
\end{eqnarray}
Well-known relations between the sides and  medians of a triangle yield
\begin{eqnarray}
&&k^2=2(a^2+c^2)-b^2,\quad l^2=2(a^2+b^2)-c^2,\quad m^2=2(b^2+c^2)-a^2,
\label{klmabc}
\end{eqnarray}
see \cite{Johnson}. The $\triangle$-robot moves due to the prescribed oscillations of the sides
\begin{eqnarray}
&&k=\breve{k}\equiv\overline{k}+\varepsilon\widetilde{k}(\tau), \
l=\breve{l}\equiv\overline{l}+\varepsilon\widetilde{l}(\tau),
\ m=\breve{m}\equiv\overline{m}+\varepsilon\widetilde{m}(\tau);\
\tau\equiv\omega t\label{constraints}
\end{eqnarray}
where $\omega=\const$ $\varepsilon=\const$; $\overline{k}$, $\overline{l}$, $\overline{m}$ are mean lengths,
and $\widetilde{k}$, $\widetilde{l}$, $\widetilde{m}$ are $2\pi$-periodic functions of $\tau$ with zero
average values. The spheres experience external friction forces $\vF^{(\nu)}=(F_1^{(\nu)},F_2^{(\nu)})$,
while the rods are so thin in comparison with $R$, that their interaction with a fluid is negligible. The
considered problem contains three characteristic lengths: the length of rods $L$, the radius of spheres $R$,
and the amplitude  of rod's oscillations $a$. The characteristic time-scale is $T$ and the characteristic
force is $F$. We have chosen
\begin{eqnarray}\label{scales}
&&3L\equiv \overline{k}+\overline{l}+\overline{m}, \ a\equiv\varepsilon L,\  T\equiv 1/\omega, \ F\equiv
6\pi\eta RL/T
\end{eqnarray}
where $\eta$ is viscosity of a fluid. Two independent small parameters are
\begin{eqnarray}\label{small-par}
&&\varepsilon\equiv a/L,\quad \delta\equiv 3R/(4L)
\end{eqnarray}
The dimensionless variables  (marked with asterisks)  are chosen as
\begin{eqnarray}
&& \vx=L\vx^*,\quad  t=T\,t^*,\quad
 F_i=F\,F_i^*;
\end{eqnarray}
Below we use only dimensionless variables, however all asterisks are omitted.

The form of a triangle is completely prescribed by three lengths $l,k,m$ or, alternatively, by three lengths
$a,b,c$ (\ref{abc-units}),(\ref{klmabc}). We choose generalised coordinates as
\begin{eqnarray}
&&\vq=(q_1,q_2,q_3,q_4,q_5,q_6)\equiv(X,Y,\theta,a,b,c)\label{q}
\end{eqnarray}
The choice $a,b,c$ as generalised coordinates drastically simplifies further calculations. The motion of the
$\triangle$-robot is described by the Lagrangian function $\mathcal{L}=\mathcal{L}(\vq,\vq_t)$, which
includes constraints (\ref{constraints}) with Lagrangian multipliers $\alpha$, $\beta$, $\gamma$
\begin{eqnarray}
&&\mathcal{L}(\vq,\vq_t)=\mathcal{K}+\alpha(a-\breve{a})+\beta(b-\breve{b})+\gamma(c-\breve{c})
\label{Lagr-contstr}
\end{eqnarray}
where subscript $t$ stands for $d/dt$, given functions $\breve{a}$, $\breve{b}$, $\breve{c}$ correspond to
$\breve{l}$, $\breve{l}$, $\breve{m}$ in (\ref{constraints}),(\ref{klmabc}),  and $\mathcal{K}$ is kinetic
energy of a robot. Lagrangian multipliers represent additional unknown functions of time. The Lagrange
equations are
\begin{eqnarray}
&&\frac{d}{dt}\frac{\partial \mathcal{L}}{\partial q_{n t}} -\frac{\partial \mathcal{L}}{\partial
q_n}=Q_n,\quad Q_n=\sum_{\nu=1}^3\sum_{j=1}^2 F_j^{(\nu)}\frac{\partial x_j^{(\nu)}}{\partial
q_n}\label{Lagr-eqns}
\end{eqnarray}
where $\vQ=(Q_1,Q_2,Q_3)$ is the generalized external force, exerted by a fluid on the $\triangle$-robot. As
one can see, we use latin subscripts ($i,k=1,2$) for cartesian components of vectors and tensors, subscript
$n$ for generalised coordinates, and subscripts (or superscripts) $\mu,\nu=1,2,3$ to identify the spheres.
The fluid flow past the $\triangle$-robot is described by the Stokes equations, where all inertial effects
are neglected. Since masses of spheres and rods are negligible, then $\mathcal{K}\equiv 0$. Hence,
(\ref{Lagr-contstr}),(\ref{Lagr-eqns}),(\ref{abc}),(\ref{abc-units}) produce the system of equations
\begin{eqnarray}
&&\vF^{(1)}+\vF^{(2)}+\vF^{(3)}=0\label{sum-forces}\\
&&\left[\va\times \vF^{(1)}+\vb\times \vF^{(2)}+\vc\times \vF^{(3)}\right]_\bot=0\label{sum-torques}\\
&&\va\cdot\vF^{(1)}=-\alpha\, a,\quad \vb\cdot\vF^{(2)}=-\beta\, b,\quad
\vc\cdot\vF^{(3)}=-\gamma\, c\label{eqns-reactions}\\
&&a=\breve{a},\quad b=\breve{b},\quad c=\breve{c}\label{eqns-constrs}
\end{eqnarray}
where the subscript $\bot$ stands for a perpendicular to $(x,y)$-plane component of a vector. One can see
that for chosen generalised coordinates the Lagrangian equations appear as zero total force
(\ref{sum-forces}) and zero total torque (\ref{sum-torques}). The explicit expressions for $\vF^{(\nu)}$ are
\begin{eqnarray}\label{forces-Stokes}
&&\vF^{(\nu)}\simeq -\vx_t^{(\nu)}+
\delta \vPi^{(\nu)},\quad
\vPi^{(\nu)}\equiv\sum_{\mu\neq\nu}\mathbb{S}^{(\nu\mu)}\vx_t^{(\mu)}\\
&&\mathbb{S}^{(\nu\mu)}=\mathbb{S}^{(\mu\nu)};\quad l^3\mathbb{S}^{(12)}=l^3 S_{ik}^{(12)}\equiv
l^2\delta_{ik}+l_i l_k,\quad
\emph{etc.}\nonumber
\end{eqnarray}
Each force $\vF^{(\nu)}$ represents the first approximation for the Stokes friction force exerted on a sphere
moving in the flow generated by other two spheres. To construct (\ref{forces-Stokes}) we use a classical
explicit formula for  fluid velocity past a moving sphere, see
\cite{Lamb,Landau,Moffatt}.
The equations (\ref{sum-forces})-(\ref{forces-Stokes}) represent a system of nine equations for nine unknown
functions of time:\ $X$, $Y$, $\theta$, $a$, $b$, $c$, $\alpha$, $\beta$, and $\gamma$. It is well known from
analytical mechanics that  Lagrangian multipliers represent the reactions of constraints; hence the equations
(\ref{eqns-reactions}) can be kept out of consideration if we are not interested in the forces exerted by
rods.

Since the explicit form of system (\ref{sum-forces})-(\ref{forces-Stokes}) is rather cumbersome, we restrict
ourselves to the special case of an equilateral mean triangle $\overline{k}=\overline{l}=\overline{m}=1$.
Using (\ref{abc}) we can rewrite the correspondent equations (\ref{sum-forces}),(\ref{sum-torques}) as
\begin{eqnarray}
&&3\vX_t=\delta\left[\mathbb{S}^{(12)}(2\vX_t-\vc_t)+\mathbb{S}^{(13)}(2\vX_t-\vb_t)+
\mathbb{S}^{(23)}(2\vX_t-\va_t)\right]\label{forces-iso}\\
&&\left[\va\times\va_t+\vb\times\vb_t+\vc\times\vc_t\right]_\bot=\delta\left[\va\times\vPi^{(1)}+
\vb\times\vPi^{(2)}+\vc\times\vPi^{(3)}\right]_\bot
\label{torq-iso}
\end{eqnarray}

\section{Two-timing method and asymptotic procedure \label{sect04}}

\subsection{Functions and notations}

The following \emph{dimensionless} notations and definitions are in use:

\noindent
(i) $s$ and $\tau$ denote slow time and fast time;  subscripts $\tau$ and $s$ stand for  related partial
derivatives.

\noindent
(ii) A dimensionless function, say $G=G(s,\tau)$, belongs to class $\cal{I}$ if $G={O}(1)$ and all  partial
$s$-, and $\tau$-derivatives of $G$ (required for our consideration) are also ${O}(1)$. In this paper all
functions belong to   class $\cal{I}$, while all small parameters appear as explicit multipliers.

\noindent
(iii) We consider only \emph{periodic in $\tau$ functions} $
\{G\in  \mathcal{P}:\quad G(s, \tau)=G(s,\tau+2\pi)\},
$ where $s$-dependence is not specified. Hence, all considered below functions belong to $\cal{P}\bigcap
\cal{I}$.

\noindent
(iv) For  arbitrary $G\in \cal{P}$ the \emph{averaging operation} is
\begin{eqnarray}
\langle {G}\,\rangle \equiv \frac{1}{2\pi}\int_{\tau_0}^{\tau_0+2\pi}
G(s, \tau)\,d\tau\equiv \overline{G}(s),\qquad\forall\ \tau_0\label{oper-1}
\end{eqnarray}

\noindent
(v)  \emph{The tilde-function} (or purely oscillating function) represents a special case of
$\cal{P}$-function with zero average $\langle\widetilde G \,\rangle =0$. The \emph{bar-function} (or
mean-function) $\overline{G}=\overline{G}(s)$ does not depend on $\tau$. A unique decomposition
$G=\overline{G}+\widetilde{G}$ is valid.

\subsection{Asymptotic procedure and successive approximations}

The introduction of  fast time variable $\tau$ and slow time variable $s$ represents a crucial step in our
asymptotic procedure. We choose $\tau=t$ and $s=\varepsilon^2 t$. This choice can be justified by the same
distinguished limit arguments as in \cite{VladimirovMHD}. Here we present this choice without proof, however
the most important part of this proof (that this choice  leads to a valid asymptotic procedure) is exposed
and exploited below. We use the chain rule
\begin{eqnarray}\label{chain}
&&d/dt=\partial/\partial\tau+\varepsilon^2\partial/\partial s
\end{eqnarray}
and then we  accept (temporarily) that $\tau$ and $s$ represent two independent variables. In further
consideration we consider series in small parameter $\varepsilon$ and restrict our attention to the terms
$O(\varepsilon^2)$. Simultaneously we keep at most linear terms in $\delta$. In our theory $\delta$ appears
not separately but as  products such as $\varepsilon^2\delta$. We do not specify the dependence of unknown
functions on $\delta$; such dependence reveals itself naturally during  calculations. Unknown functions are
taken as regular series
\begin{eqnarray}\label{x-f-ser}
&&\vX(\tau,s)=\vX_0(\tau,s)+\varepsilon \vX_1(\tau,s)+\varepsilon^2 \vX_2(\tau,s)+\dots
\end{eqnarray}
and similar series for $\theta$, $a$, $b$, $c$, $\alpha$, $\beta$, and $\gamma$.  We take
$$
\widetilde{\vX}_0\equiv 0\  \text{and}\ \widetilde{\theta}_0\equiv 0,
\quad\text{while}\quad \overline{\vX}_0\neq 0\ \text{and}\ \overline{\theta}_0\neq 0
$$
which express the basic property of our solutions: long distances of self-swimming and large angles of
rotations are caused by small oscillations. We  also have accepted
$\overline{a}_0=\overline{b}_0=\overline{c}_0=1/\sqrt{3}$ for an equilateral triangle. After the application
of (\ref{chain}) to (\ref{x-f-ser}) we have
\begin{eqnarray}\label{x-f-ser1}
&&\vX_t=\varepsilon \widetilde{\vX}_{1\tau}+\varepsilon^2
(\widetilde{\vX}_{2\tau}+\overline{\vX}_{0s})+O(\varepsilon^2)
\end{eqnarray}
and similar expression for $\theta$. In calculations below all the bar functions belong to the zero
approximation and all the tilde-functions belong to the first approximation, therefore we omit the related
subscripts.

The successive approximations of equations (\ref{forces-iso}),(\ref{torq-iso}) yield:

\noindent \emph{Terms} $O(\varepsilon^0)$ give identities $0=0$.

\noindent  \emph{Terms}  $O(\varepsilon^1\delta^0)$ lead to
\begin{eqnarray}
&&\widetilde{\vX}_{\tau}=0,\quad
3\widetilde{\theta}_{\tau}+2\widetilde{\psi}_{\tau}+\widetilde{\chi}_{\tau}=0
\label{first-order}
\end{eqnarray}
where the second equation follows from
\begin{eqnarray}
&&\left[\overline{\va}\times\widetilde{\va}_{\tau}+\overline{\vb}\times\widetilde{\vb}_{\tau}+
\overline{\vc}\times\widetilde{\vc}\right]_\bot=0
\label{torq-iso-1}
\end{eqnarray}
and definitions (\ref{abc-units}). The integration of (\ref{first-order}) in the class of periodic functions
yields
\begin{eqnarray}
&&\widetilde{\vX}=0,\quad 3\widetilde{\theta}+2\widetilde{\psi}+\widetilde{\chi}=0
\label{first-order-1}
\end{eqnarray}
From the law of cosines one can derive that $
\widetilde{\psi}=\widetilde{a}+\widetilde{b}-2\widetilde{c}$ and
$\widetilde{\chi}=\widetilde{b}+\widetilde{c}-2\widetilde{a}$, then (\ref{first-order-1}) gives
\begin{eqnarray}
&&\widetilde{\theta}=\widetilde{c}-\widetilde{b}.
\label{theta-1}
\end{eqnarray}
\noindent \emph{Terms} $O(\varepsilon\delta)$: These terms do not vanish and can be easily calculated.
However, (as one can see below) they do not participate in the leading terms of average motion that appear in
the order $O(\varepsilon^2)$.

\noindent \emph{Terms} $O(\varepsilon^2)$: Eqns.(\ref{forces-iso}),(\ref{torq-iso}) give
\begin{eqnarray}\label{eqn-X-2}
&& 3\overline{\vX}_{s}=\delta\langle \widetilde{\mathbb{S}}_\tau^{(12)}\widetilde{\vc}_\tau+
\widetilde{\mathbb{S}}_\tau^{(13)}\widetilde{\vb}_\tau+
\widetilde{\mathbb{S}}_\tau^{(23)}\widetilde{\va}_\tau\rangle\\
&&
\left[\overline{\va}\times\overline{\va}_{s}+\overline{\vb}\times\overline{\vb}_{s}+
\overline{\vc}\times\overline{\vc}_{s}\right]_\bot=
\langle\left[\widetilde{\va}\times\widetilde{\va}_{\tau}+\widetilde{\vb}\times\widetilde{\vb}_{\tau}+
\widetilde{\vc}\times\widetilde{\vc}_{\tau}\right]_\bot\rangle
\label{eqn-theta-2}
\end{eqnarray}
From these equations one can already see that the leading terms $\overline{\vX}_{0s}=O(\delta)$ and
$\overline{\theta}_{0s}=O(1)$.  In further calculations all  tilde-functions, say,
\begin{eqnarray}
&&\widetilde{S}_{ik\tau}^{(12)}=-(\delta_{ik}+3\overline{l}_{i}\overline{l}_{k})\widetilde{l}_\tau+
\overline{l}_{i}\widetilde{l}_{k\tau}+\overline{l}_{k}\widetilde{l}_{i\tau},\quad \emph{etc.}
\label{S-tau}
\end{eqnarray}
should be expressed in terms of $\va$, $\vb$, $\vc$ with the use of (\ref{abc}),(\ref{abc-units}). Then
straightforward but rather cumbersome transformations of (\ref{eqn-X-2}) yield
\begin{eqnarray}
&& 3\overline{\vX}_{s}/\delta=\langle \widetilde{a}\widetilde{b}_\tau\rangle \widehat{\vc}'
+\langle\widetilde{c}\widetilde{a}_\tau\rangle \widehat{\vb}'+
\langle\widetilde{b}\widetilde{c}_\tau\rangle \widehat{\va}'\label{eqn-X-3}\\
&& =\frac{\sqrt{3}}{2}\langle \widetilde{a}\widetilde{b}_\tau-
\widetilde{c}\widetilde{a}_\tau\rangle \widehat{\va}+
\frac{1}{2}\langle 2\widetilde{b}\widetilde{c}_\tau-\widetilde{a}\widetilde{b}_\tau
-\widetilde{c}\widetilde{a}_\tau\rangle \widehat{\va}'\nonumber\\
&&
\overline{\theta}_s=\frac{4}{\sqrt{3}}\langle \widetilde{a}\widetilde{c}_\tau+\widetilde{b}\widetilde{c}_\tau+
\widetilde{c}_\tau \widetilde{b}\rangle
\label{eqn-theta-3}
\end{eqnarray}
where unit vectors $\widehat{\va}'$,  $\widehat{\vb}'$, and $\widehat{\vc}'$ correspond to
$\theta$-derivatives of $\widehat{\va}$,  $\widehat{\vb}$, and $\widehat{\vc}$, say,
\begin{eqnarray}\label{abc-units-prime}
&&\widehat{{\va}}'=
\left(
\begin{array}{c}
    -\sin\overline{\theta}  \\
     \cos\overline{\theta}
  \end{array}
\right),\quad
\widehat{\vb}=\frac{1}{2}(-\widehat{\va}+\sqrt{3}\widehat{\va}'),\quad
\widehat{\vb}'=\frac{1}{2}(-\sqrt{3}\widehat{\va}-\widehat{\va}'),\quad \emph{etc.}
\nonumber
\end{eqnarray}
One can check that expressions (\ref{eqn-X-3}),(\ref{eqn-theta-3}) possess a relabeling invariance
$(\va,\vb,\vc)\mapsto (\vb,\vc,\va)$, \emph{etc}. The final expressions are
\begin{eqnarray}
&& \overline{\vX}_{s} = \frac{\delta}{9}(A \widehat{\va}+B\widehat{\va}'),\quad
\overline{\theta}_s=D;\label{eqn-X-theta}\\
&&A\equiv\frac{\sqrt{3}}{2}
\langle (\widetilde{l}+\widetilde{k})\,\widetilde{m}_\tau
\rangle,\quad
B\equiv \frac{1}{2}\langle 2\widetilde{k}\widetilde{l}_\tau-\widetilde{l}\widetilde{m}_\tau
-\widetilde{m}\widetilde{k}_\tau\rangle\label{A-B}\\
&&D\equiv-\frac{4}{3\sqrt{3}}\langle
\widetilde{l}\,\widetilde{m}_\tau+\widetilde{k}\,\widetilde{l}_\tau+
\widetilde{m}\, \widetilde{k}_\tau\rangle\label{D}
\end{eqnarray}
with constants $A$, $A'$, and $\Omega$. It is also useful to exclude $m$ by introducing the angle $\varphi$
between the sides $k$ and $l$. The law of cosines gives
$\widetilde{m}=(\widetilde{k}+\widetilde{l}+\sqrt{3}\widetilde{\varphi}/2)/2$; its substitution to
(\ref{A-B}),(\ref{D}) yields
\begin{eqnarray}
&&A\equiv\frac{3}{8}
\langle (\widetilde{l}+\widetilde{k})\,\widetilde{\varphi}_\tau
\rangle,\quad
B\equiv \frac{1}{2}\langle
-3\widetilde{k}\widetilde{l}_\tau+\frac{\sqrt{3}}{4}(\widetilde{l}-\widetilde{k})\widetilde{\varphi}_\tau
\rangle,\quad D\equiv-\frac{1}{3}\langle (\widetilde{l}-\widetilde{k})\widetilde{\varphi}_\tau\rangle\label{D1}
\end{eqnarray}
Both equations (\ref{eqn-X-theta}) can be immediately integrated as
\begin{eqnarray}
&&\overline{X}-X_0=\frac{\delta}{9D}(A\sin\overline{\theta}+B\cos\overline{\theta}),\quad
\overline{Y}-Y_0=\frac{\delta}{9D}(-A\cos\overline{\theta}+B\sin\overline{\theta})\label{circle}\\
&&\overline{\theta}=D s+ \theta_0\nonumber
\end{eqnarray}
with constants of integration $X_0$, $Y_0$, and $\theta_0$. Finally, we should recall that $s$ is slow time
variable, hence physical translational velocity and angular velocity are
\begin{eqnarray}
&& \overline{\vV}\equiv\overline{\vX}_{t} = \frac{\delta\varepsilon^2}{9}(A
\widehat{\va}+B\widehat{\va}'),\quad
\Omega\equiv\overline{\theta}_t=\varepsilon^2D \label{eqn-X-final}
\end{eqnarray}
For dimensional velocities $\overline{\vV}$ and $\Omega$ one has to add  multipliers $L\omega$ and $\omega$
correspondingly (see (\ref{scales})). It is noticeable that $\overline{\vV}$ and $\Omega$ have different
amplitudes: $\overline{\vV}=O(\delta\varepsilon^2)$, but $\Omega=O(\varepsilon^2)$. It corresponds to an
important physical fact:   translational self-propulsion of a $\triangle$-robot takes place due to
hydrodynamic interactions between spheres, while rotational self-propulsion does exist without taking such
interactions into account. Eqns.(\ref{circle}) show that the $\triangle$-robot rotates with  constant angular
velocity $\Omega$ around its own centroid and simultaneously moves with the same angular velocity $\Omega$ in
a circular orbit of radius $\delta\sqrt{A^2+B^2}/9\Omega$ and centered at the point $(X_0,Y_0)$, which can be
defined from the initial data.

\section{Examples of rectilinear self-propulsion}

The condition for motion (\ref{eqn-X-theta}) to be purely translational is $\Omega= 0$; it could be really
useful for practical applications. The related absolute value of translational velocity is
\begin{eqnarray}
|\overline{\vV}|=\delta\varepsilon^2\sqrt{A^2+B^2}/9\label{V}
\end{eqnarray}
Hence, the equation
\begin{eqnarray}
\langle\widetilde{l}\,\widetilde{m}_\tau+\widetilde{k}\,\widetilde{l}_\tau+
\widetilde{m}\, \widetilde{k}_\tau\rangle=0\quad \text{or}\quad
\langle (\widetilde{l}-\widetilde{k})\widetilde{\varphi}_\tau\rangle=0\label{straight-cond}
\end{eqnarray}
provides options for purely translational (rectilinear) motion. The simplest cases correspond to  isosceles
oscillations, when $\widetilde{k}=\widetilde{l}\neq\widetilde{m}$ or
$\widetilde{k}=\widetilde{m}\neq\widetilde{l}$ or $\widetilde{l}=\widetilde{m}\neq\widetilde{k}$. For
example, $\widetilde{k}=\widetilde{l}\neq\widetilde{m}$ gives
$$
\Omega=0,\quad A=\sqrt{3}\langle \widetilde{k}\widetilde{m}_\tau\rangle=
3\langle \widetilde{l}\widetilde{\varphi}_\tau\rangle/4,\quad B=0
$$
which coincides with the translational self-swimming of a $V$-robot by \cite{VladimirovX4}. However, in
\cite{VladimirovX4} the mean angle between the arms of a $V$-robot is arbitrary,
while everywhere in this paper it is $\pi/3$. Hence, the results of
\cite{VladimirovX4} are complementary to the study done in this paper.
It is also worth to emphasise that variables and the method of calculation in
\cite{VladimirovX4} are different, hence the obtaining of the same result strongly suggests that
the calculations of this paper do not contain mistakes.

Another case of rectilinear motion is $\varphi=0$, which corresponds to a fixed angle between the sides $k$
and $l$, while $\widetilde{k}\neq 0$ and $\widetilde{l}\neq 0$. From (\ref{D1}) one can see that in this case
$A=0$, $D=0$, and $B=3\langle
\widetilde{l}\widetilde{k}_\tau\rangle/2$. It gives  velocity
$
\vV={\delta\varepsilon^2}\langle \widetilde{l}\widetilde{k}_\tau\rangle\widehat{\va}'/6
$, which shows that the $\triangle$-robot is not rotating but propagating in  direction  $\overline{\vm}$.
This example can be seen as  generalization of the micro-swimmer by \cite{Golestanian} for a fixed angle
$\pi/3$ between the arms. Similarly, the examples of pure rotation (without translation) can be build.

\section{Discussion}

(i)  We have derived that for any oscillations of three sides of the considered  $\triangle$-robot the
averaged  paths of each sphere represent circles with centers that move in circles (\ref{circle}). That is,
the paths of each sphere are epicycloids. Following  other examples in fluid dynamics (see
\cite{Bennett}) one can call these paths \emph{Ptolemaic solutions}.

(ii) The magnitude $O(\varepsilon^2\delta)$ of translational velocity in terms of small parameters
(\ref{eqn-X-final}),(\ref{V}) is the same as the result by
\cite{Golestanian, Golestanian1,VladimirovX3} for linear micro-robots.  At the same time, our choice of
slow time $s=\varepsilon^2 t$ (\ref{chain}) agrees with classical studies of self-propulsion for low Reynolds
numbers, see \cite{Taylor, Blake, Childress}, as well as the geometric studies of \cite{Wilczek}.

(iii) The magnitude $O(\varepsilon^2)$ for angular velocity (\ref{eqn-X-final}) represents a novel result. It
corresponds to an important physical fact: rotational self-propulsion does exist without taking hydrodynamic
interactions between spheres into account; it caused by  standard Stokes' drag force in an infinite fluid and
the reactions of constraints.

(iv) We have built an asymptotic procedure with two small parameters: $\varepsilon\to 0$ and $\delta\to 0$.
Such a setting usually requires the consideration of different asymptotic paths on the plane
$(\varepsilon,\delta)$ when, say $\delta=\delta(\varepsilon)$. In our case we can avoid it, since $\delta$
does not appear separately,  but only in combinations such as $\varepsilon^2\delta$.

(v) The mathematical justification of the presented results can be performed by the estimation of an error in
the original equation, as in  \cite{VladimirovX1,VladimirovX2}. It is also possible to derive the higher
approximations of $\overline{\vV}$ and $\Omega$, as it has been done by
\cite{VladimirovX1,VladimirovX2} for different cases. These approximations can be useful to study
the cases when  self-propulsion in the main order vanishes.

(vi) In order to compare velocities of micro-robots and micro-organisms we use dimensional variables, where
$\overline{V}\sim \omega L
\varepsilon^2\delta$; it  shows that a $\triangle$-robot can move itself with the speed  about $10\%$
of its own size per second (we have taken $\varepsilon=\delta=0.2$ and $\omega=30 s^{-1}$; the value of
$\omega$ can be found in \cite{PedKes, VladPedl, Pedley, Polin}. From these papers we also can see that this
estimation of self-propulsion velocity is about $10$ times lower than a similar value for natural
micro-swimmers.

\begin{acknowledgments}
The author is grateful to Profs. M. Bees,  A. Gilbert, K.I. Ilin, and J. Pitchford for useful discussions.
\end{acknowledgments}

\end{document}